\documentclass[10pt,letterpaper,twocolumn]{article} 

\usepackage{ol2}
\usepackage{hyperref}
\usepackage{amsmath}
\usepackage{graphicx}

\begin{document}

\twocolumn[ 

\title{Laser cooling with a single laser beam and a planar diffractor}


\author{Matthieu Vangeleyn, Paul F. Griffin, Erling Riis and Aidan S. Arnold$^*$}

\address{Department of Physics, SUPA, University of Strathclyde, Glasgow G4 0NG, UK\\
$^*$Corresponding author: a.arnold@phys.strath.ac.uk }

\begin{abstract}
A planar triplet of diffraction gratings is used to transform a single laser beam into a four-beam tetrahedral magneto-optical trap. This `flat'
pyramid diffractor geometry is ideal for future microfabrication. We demonstrate the technique by trapping and subsequently sub-Doppler cooling
$^{87}$Rb atoms to $30\,\mu$K.
\end{abstract}


 ] 


A magneto-optical trap (MOT) \cite{raab} is the starting point for the vast majority of cold and ultracold atomic physics experiments. Atoms are
trapped and cooled to sub-milliKelvin temperatures using light scattering modified by the Zeeman and Doppler effects, respectively. MOTs are
typically formed at the center of a spherical quadrupole magnetic field, in the overlap region of six (or less commonly four \cite{shimizu})
appropriately polarised red-detuned laser beams. The original pyramid MOT (PMOT) \cite{Jhe}, utilising a square-based pyramidal reflector with
$90\,^{\circ}$ apex angle between opposite sides, was devised as a means to turn a single laser beam into the six appropriately polarised beams
required for a MOT. The PMOT simplifies optical alignment, saves a large number of optical components, and can also be modified to produce a beam
source of cold atoms \cite{foot}. The original PMOT has since been used to make a compact gravimeter \cite{Landragin}, and a millimetre scale chip
trap \cite{ed}.

Recently we demonstrated a new kind of pyramid MOT, based on a four-beam tetrahedral geometry \cite{vangeleyn} originating from a single beam
interacting with a triangular pyramid reflector. This geometry has many advantages over the original design: MOT formation outside the pyramid is
possible which simplifies optical access to the atoms and the apex region of the pyramid is non-critical (the latter feature is also present in the
PMOT design in Ref.~\cite{zhan}). The apex and mirror edges in the original PMOT will generate diffraction and incorrect apex angle will also
generate intensity irregularities in the doubly-reflected beams counterpropagating with the input beam. As these irregularities pass directly through
the MOT, they can hinder further cooling in optical molasses. Moreover although sub-Doppler cooling is possible with small atom number
\cite{Landragin}, for larger PMOTs the counterpropagating beam will contain a shadow of the atoms from the input beam, creating an intensity
imbalance that will hinder molasses \cite{lett}. This problem is obviated in the tetrahedral PMOT.

\begin{figure}[!b]
\centerline{\includegraphics[width=.76\columnwidth]{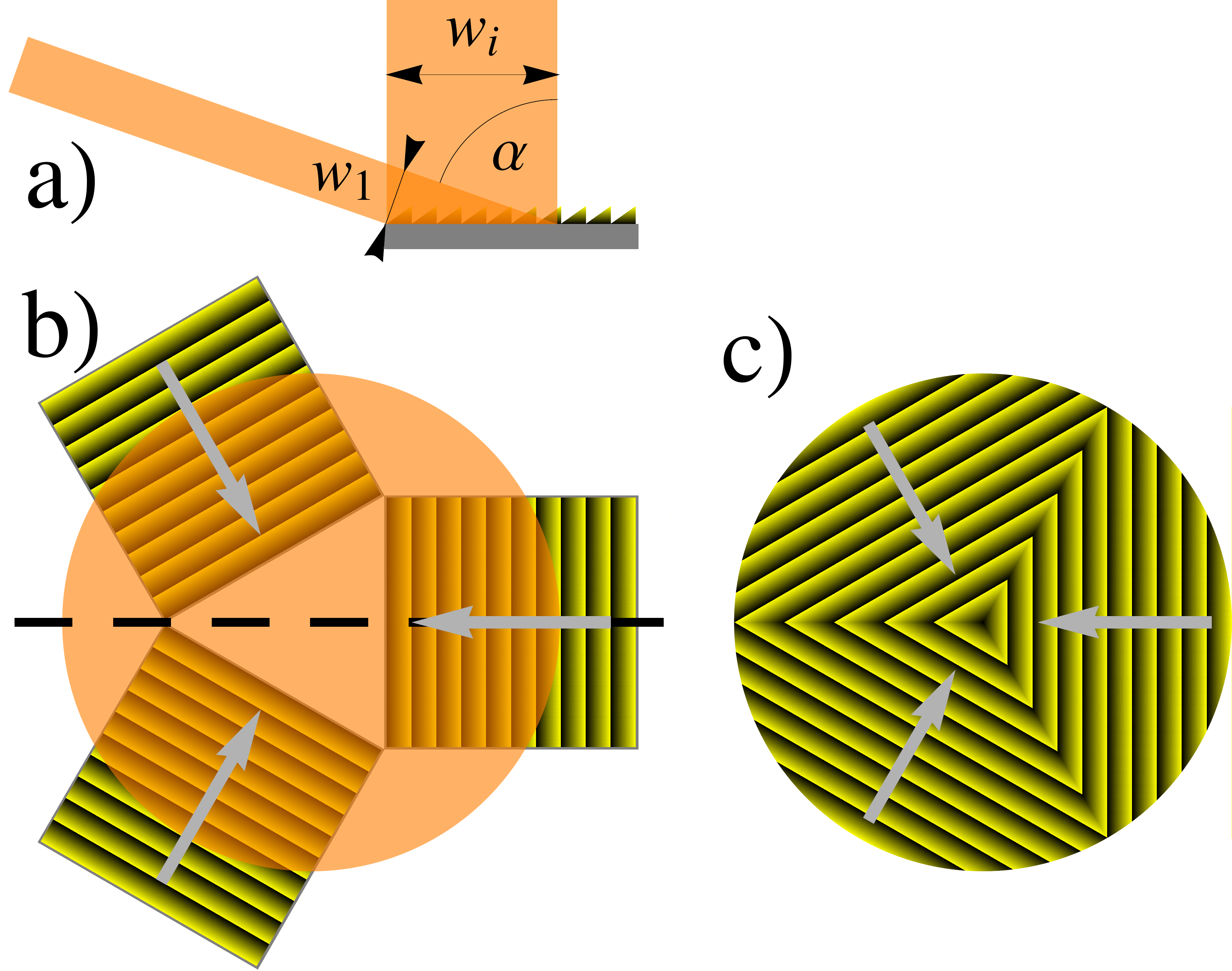}}
 \caption{a) Side view of a first-order diffracted beam, illustrating geometric beam compression. b) Top view of three gratings and the circular
 vertical incident beam. The arrows show the blaze direction. c) A design for improved use of the laser beam area. Tomographic movies of the beam
 overlap volume for b) and c) are shown in Media 1, Media 2.} \label{figGratingSchem}
\end{figure}

In this letter we have experimentally realized our proposal to extend the tetrahedral PMOT \cite{vangeleyn} to a `flat' geometry using diffraction
gratings. The grating magneto-optical trap (GMOT) has a very similar working principle to the tetrahedral PMOT, and its properties are again largely
the result of intensity balance and polarisation decomposition \cite{vangeleyn}. One major difference is due to the fact that gratings spatially
compress beams with a corresponding intensity increase (Fig.~\ref{figGratingSchem} a). The relationship between the intensities $I_\textrm{i}$ and
$I_1$ of the vertical incident beam and the first-order diffracted beam, respectively, is determined by the corresponding beam widths $w_\textrm{i}$
and $w_1$, and the first order diffraction efficiency $R_1$:
\begin{equation}
    I_\textrm{1}/I_\textrm{i}=R_1\, w_\textrm{i}/w_1= R_1 \sec{\alpha},
    \label{eqintrel}
\end{equation}
where the Bragg condition yields the first order diffraction angle $\alpha=\arcsin{(\lambda/d)}$ for light with wavelength $\lambda$ normally
incident on a grating with groove spacing $d$. Note the relation $\alpha=2\theta$ allows direct comparison with mirror declination angle $\theta$ in
our previous work \cite{vangeleyn}.

The condition for balanced optical molasses from beams with intensities $I_j$ and wavevectors $\textbf{k}_j$ is:
\begin{equation}
    \sum{I_j \textbf{k}_j=\textbf{0}}.
    \label{eqoptmol}
\end{equation}
If we consider the configuration depicted in Fig.~\ref{figGratingSchem} b), where all but one beam are provided by diffraction from $n$ identical
gratings, then Eq.~(\ref{eqoptmol}) is \textit{radially} always satisfied, given the symmetry of the problem. Substituting Eq.~(\ref{eqintrel}) into
Eq.~(\ref{eqoptmol}) and projecting onto the vertical axis yields the very simple condition for balanced optical molasses:
\begin{equation}
    R_1=1/n,
    \label{eqmolcond}
\end{equation}
which is completely independent of diffraction angle and hence grating period. For three beams this corresponds to first order diffraction efficiency
of $R_1=1/3.$

We consider only gratings for which second order diffraction is absent (i.e.\ first-order angles $\alpha>30^{\circ}$), as they are simpler to model
and additionally small $\alpha=2\theta$ leads to drastically reduced trapping and cooling properties \cite{vangeleyn}. To create the maximum trap
volume for a given beam size, the geometry in Fig.~\ref{figGratingSchem}c) could be used. Unlike the system corresponding to our experimental
realization (Fig.~\ref{figGratingSchem}b) the zeroth grating order (a retroreflection with efficiency $R_0$) has to be considered as it is present in
the beam overlap volume. This modifies the balanced molasses condition Eq.~(\ref{eqmolcond}) to $R_1=(1-R_0)/n,$ and all cooling forces are reduced
by the factor $1-R_0.$ The effect on the vertical trapping force depends on the relative zeroth order reflection phase shift between S and P
polarizations, tending only to improve with non-zero relative phase.

A critical point in the achievement of a tetrahedral magneto-optical trap is the circular polarization of the first order diffracted beams. Grating
efficiency is usually specified in terms of S and P polarization, and can vary dramatically with wavelength and polarisation. However, the difference
in phase accumulation between S and P components $\phi_\textrm{SP}$ also has to be taken into consideration. For our configuration, optimal cooling
and trapping is achieved when the handedness (direction of circular polarization relative to beam propagation) of the incident vertical beam is
reversed \cite{vangeleyn} and the total power drops by a factor 3 (i.e.\ all beams have equal intensity).

\begin{figure}[!t]
\centerline{\includegraphics[width=.73\columnwidth]{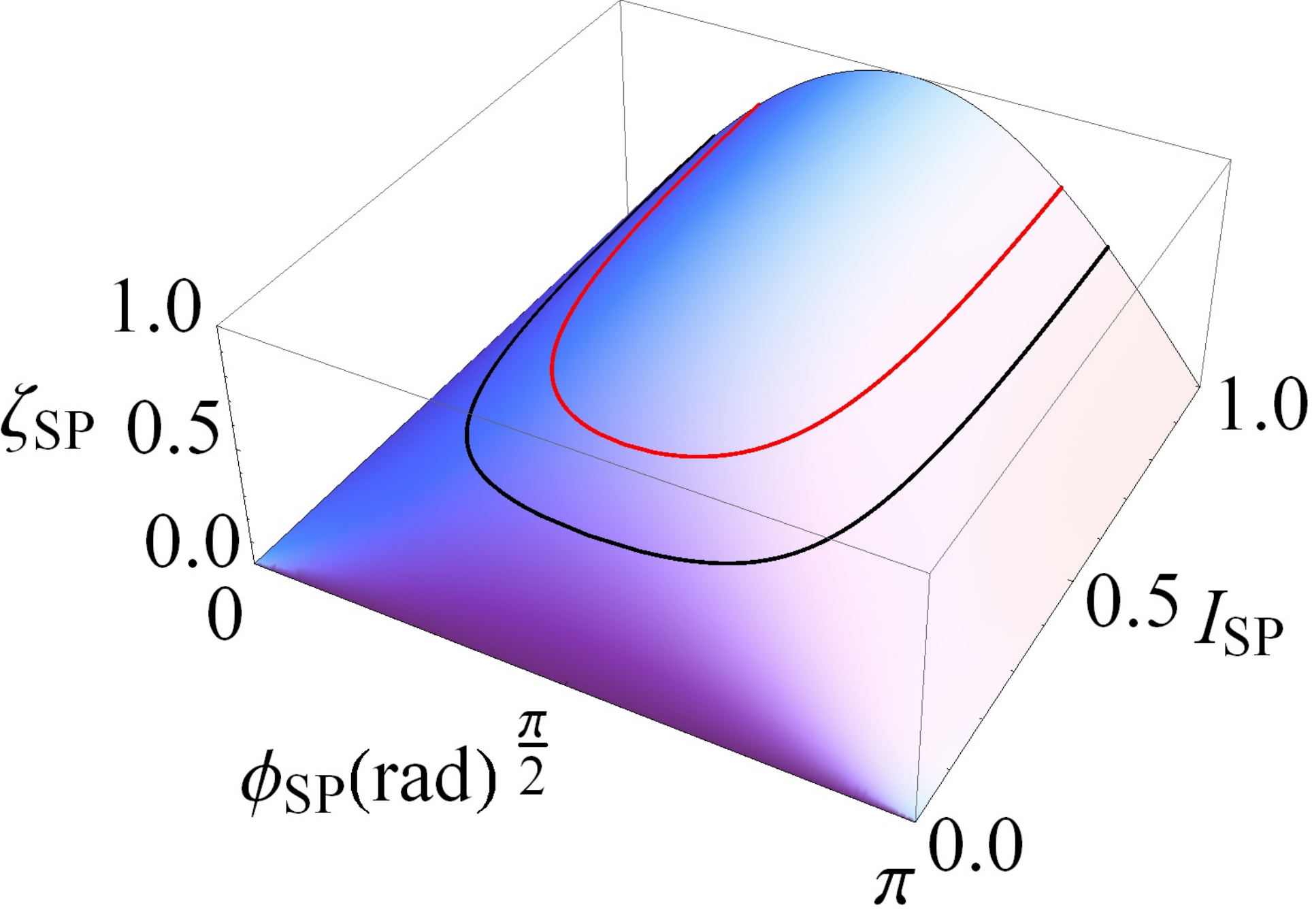}}
 \caption{Relative grating MOT radial trapping strength compared to the tetrahedral PMOT
\cite{vangeleyn}, $\eta_\textrm{SP},$ as a function of the intensity ratio $I_\textrm{SP}$ and relative phase $\phi_\textrm{SP}$ between the S and P
components of the first order beams at the MOT location. The black and red curves indicate 50\% and 70\% trapping reduction, respectively.}
\label{figth}
\end{figure}

\begin{figure}[!t]
 \centerline{\includegraphics[width=.66\columnwidth]{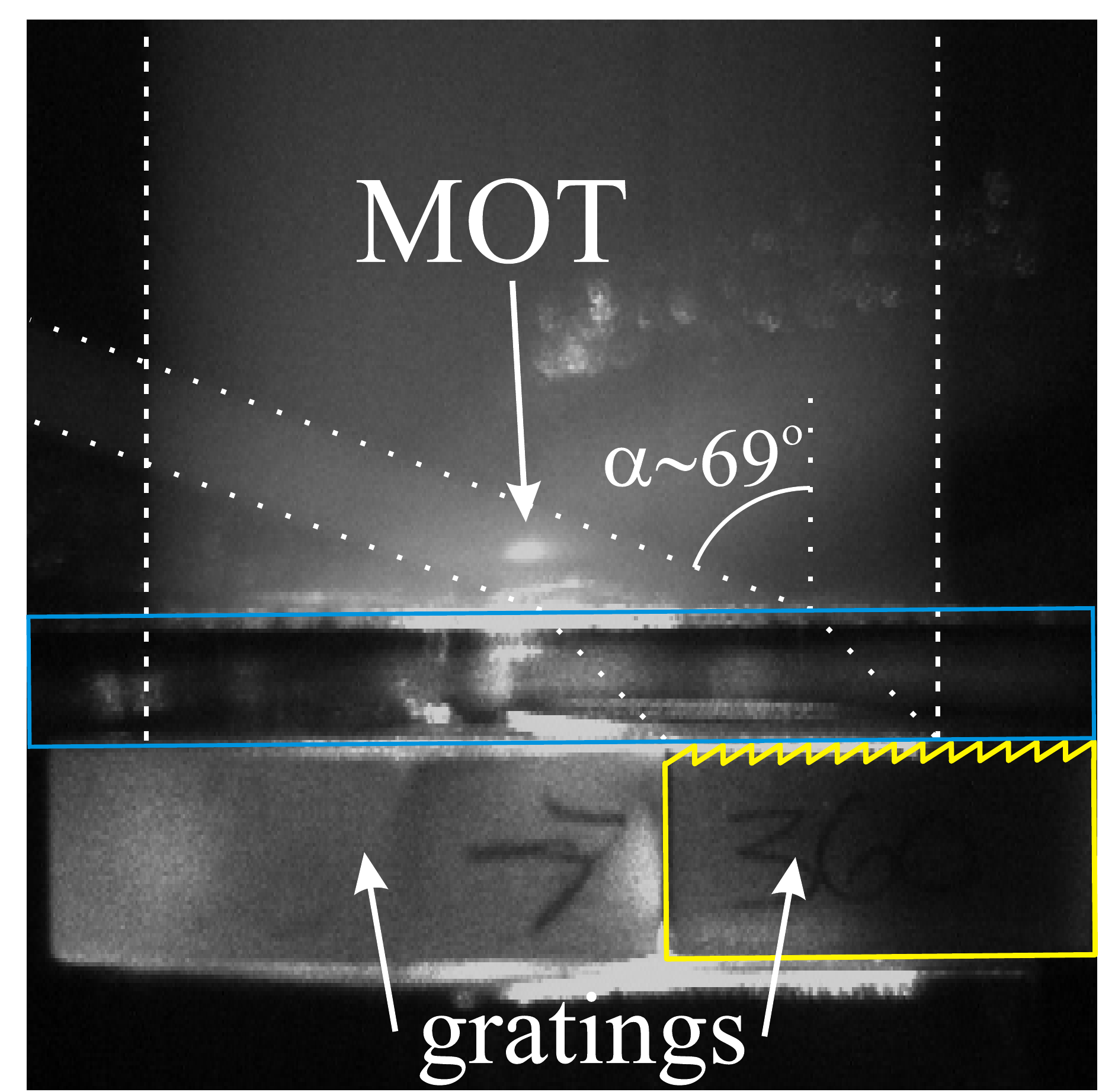}}
 \caption{Photograph of the experimental setup from the side. The MOT forms in the overlap of the downward incident beam (dashed white
lines) and the three first-order grating beams. The path of the diffracted beam (dotted white lines) from one grating (yellow schematic) refracts
through the $3\,$mm thick quartz vacuum cell (blue), then propagates at the expected $\alpha\sim69^{\circ}$.} \label{figGratingPhoto}
\end{figure}

One can show that the radial trapping constant of the GMOT, relative to the tetrahedral PMOT is reduced by a correction factor
\begin{equation}\zeta_\textrm{SP}=2 \sqrt{I_\textrm{SP}} \sin \phi_\textrm{SP}/(1+I_\textrm{SP})\label{ratio}\end{equation} if the S and P linear components of the first
order grating beams have an intensity ratio of $I_\textrm{SP}$ and a relative phase shift of $\phi_\textrm{SP}$ respectively. The effect of relative
S:P intensity ratio and phase is shown in Fig.~\ref{figth} and is surprisingly robust. For an ideal $\pi/2$ phase shift between S and P even an
intensity ratio $I_\textrm{SP}\sim 0.07$ still yields $\sim 1/2$ the trapping strength (black curve in Fig.~\ref{figth}). We measured the efficiency
of our gratings for a circularly polarized incident beam to be $R_1=45.3\%$, of which $90\,\%$ has the correct circular handedness. If we consider
the ensemble made of the grating and the quartz plate of the vacuum chamber (Fig.~\ref{figGratingPhoto}), overall diffraction efficiency drops to a
near-optimal $R_1=32.2\,\%$, with $85\,\%$ having the correct handedness, i.e.\ $\zeta_\textrm{SP}=70\%$ (red curve in Fig.~\ref{figth}).

In the experiment, trapping and repumping light are provided by two independent external cavity diode lasers \cite{diode}. The lasers are overlapped
and then spatially filtered by a $30\,\mu$m pinhole to remove rapid spatial intensity variation before and after diffraction from the gratings, which
degrades the trap loading and can prevent sub-Doppler molasses. The beam is also over-expanded such that the intensity profile is as flat as possible
within the $23\,$mm diameter apertured laser beam, to reduce intensity gradients in the three diffracted beams. A quarter-wave plate changes the
polarization to circular just before the vacuum chamber. We used inexpensive Edmund Optics gratings NT43-752. These 1200 grooves/mm gratings deflect
a normally incident $780\,$nm beam at an angle $\approx 69.4^\circ$, close to the ideal tetrahedron PMOT angle ($\arccos(1/3)\approx 70.5^\circ$).
This yields maximum trapping and cooling \cite{vangeleyn} albeit with a decreased capture volume using the grating geometry.

The grating triplet is positioned below the glass vacuum cell, the gratings forming a triangle with the blazed direction pointing towards the center
(Fig. \ref{figGratingSchem} a). The gratings have dimensions $12.7\,$mm by $12.7\,$mm and thus are not completely illuminated by the $23\,$mm
diameter laser beam. After diffraction, the beams are vertically squeezed to about $w_1=2.7\,$mm, due to compression on the gratings, and the overlap
region, where atoms can be trapped, is approximately a flattened rhombohedron \cite{vangeleyn}. The overlap volume is $\sim60\,$mm$^3$ and entirely
above the $3\,$mm thick quartz vacuum window (Fig.~\ref{figGratingPhoto}). By deliberately tilting the gratings beyond the flat geometry to increase
the beam overlap region, we found we could collect more atoms, indicating that the atom number is indeed overlap-volume limited. Ideally gratings
with a longer period could be used, however for commercial gratings the variety in blaze angle (and hence polarization-dependent diffraction
efficiency) is limited unless the spatial period is a multiple of $600\,$grooves/mm.


With intensities of $1.3\,$mW/cm$^2$ in both the vertical and diffracted beams, a magnetic field gradient of $17\,$G/cm and $7\,$MHz red-detuning, we
trap 10$^5$~$^{87}$Rb atoms in our GMOT (Fig.~\ref{figGratingPhoto}), consistent with the beam overlap volume reduction from our previous work
\cite{vangeleyn}. For our experimental parameters the atom number should only enter the volume-squared scaling regime \cite{ed} for beam diameters
less than $2\,$mm.

\begin{figure}[!t]
 \centerline{\includegraphics[width=.8\columnwidth]{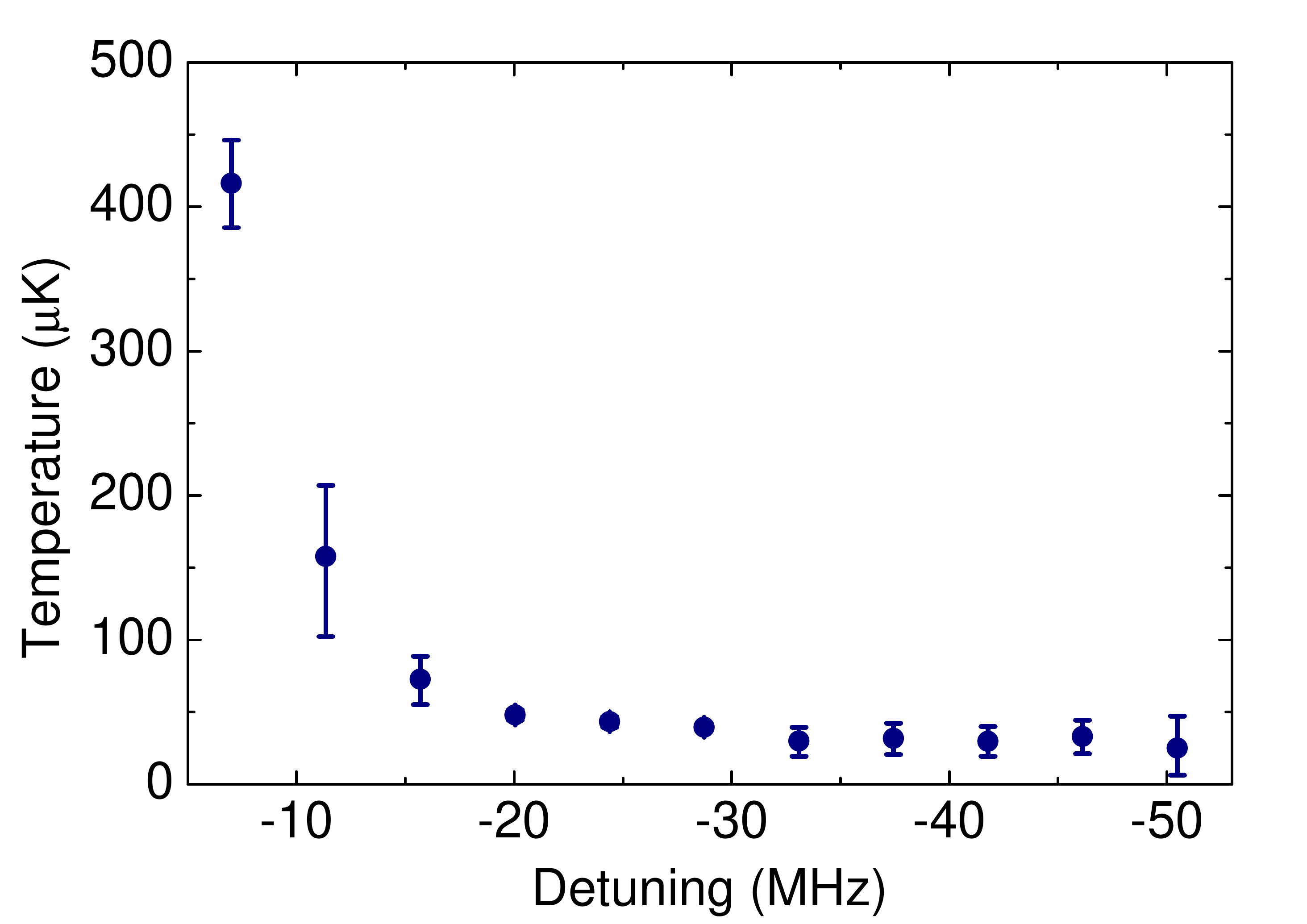}}
\caption{Temperature in the MOT as a function of the $20\,$ms extra detuning. For red-detuning jumps
 larger than $30\,$MHz, the temperature drops to $30\pm10\,\mu$K.} \label{figGraphs}
\end{figure}

After initial MOT loading, the light frequency is further red-detuned for $20\,$ms to achieve sub-Doppler $(<140\,\mu$K) MOT temperatures. The cloud
temperature is measured using the sizes of background-subtracted fluorescence images after 0 and $10\,$ms time of flight. Fig.~\ref{figGraphs} shows
the evolution of temperature as a function of extra in-MOT detuning, reaching significantly sub-Doppler temperatures of $30\pm10\,\mu$K for
red-detunings $>30\,$MHz. For a fixed $30\,$MHz extra red-detuning we investigated optical molasses formation by reducing the MOT magnetic field
gradient to a fixed value in the range $0-17\,$G/cm for the last $10\,$ms of the $20\,$ms in-MOT cooling phase. Although the temperature remains
approximately constant, as the final magnetic field gradient reaches zero the $1/e$ diameter of the Gaussian cloud prior to time of flight imaging
reaches a size $(\sim4\,$mm) comparable to the beam overlap volume.

It appears that deeper molasses cooling is largely prevented by spatial intensity variation in our small beam overlap volume, particularly near the
edge of the cloud. Under optimal conditions we have seen preliminary evidence for optical molasses, however future experiments would be better
performed in an optimized setup -- with larger beam overlap and less dramatic diffracted beam compression. Both goals can be achieved using
diffraction gratings with a larger groove period. A particularly appealing aspect of the GMOT is that it lends itself to custom microfabricated
planar optical elements, with arbitrary groove spacing in a stand-alone planar element.

In conclusion we have demonstrated a pyramid magneto-optical trap with `flat' optics, extending our work on the tetrahedral pyramid MOT
\cite{vangeleyn}. A single beam is split into three new beams by a planar diffractor. This diffractor is well-suited to microfabrication, as
technically challenging deep etching is not required. Additionally, MOT formation above the plane of the gratings has clear advantages for detection
and further manipulation of the atoms. One can envisage applications in portable MOT-based devices. Moreover, we have demonstrated sub-Doppler
temperatures in our grating MOT, and like the tetrahedral PMOT, sub-Doppler optical molasses should be achievable even with large atom number,
suitable for applications requiring Bose-Einstein condensation formation.

We are grateful for stimulating discussions with Joseph Cotter and Ed Hinds. PFG received support from the RSE/Scottish Government Marie Curie
Personal Research fellowship program.

\bibliographystyle{osajnl}

\end{document}